\documentstyle[epsf]{aipproc}
\def\pds{{\it PDS\/}}   \def\B{{\it BeppoSAX\/}}
\begin{document}

\title{Initial Results from the High Energy Experiment \pds\ aboard \B}

\author{F. Frontera$^{1,2}$, D. Dal Fiume$^1$, E. Costa$^3$, M. Feroci$^3$, \\
M. Orlandini$^1$, L. Nicastro$^1$, E. Palazzi$^1$, \\
G. Zavattini$^2$ and P. Giommi$^4$}
\address{$^1$Istituto Tecnologie e Studio Radiazioni Extraterrestri, CNR,
40126 Bologna, Italy\\
$^2$Dipartimento di Fisica, Universit\`a di Ferrara, 44100 Ferrara, Italy\\
$^3$Istituto Astrofisica Spaziale, CNR, 00144 Frascati\\
$^4$\B\ Science data Center, Roma, Italy}

\maketitle

\begin{abstract}
The high energy experiment \pds\ is one of the Narrow Field Instruments aboard
the X--ray astronomy satellite \B. It covers the energy band from 15 to 300
keV. Here we report results on its in-flight performance and observations
of galactic and extragalactic X--ray sources obtained during the
Science Verification Phase of the satellite: in particular Crab, Cen~X--3,
4U1626--67 and PKS2155--305.
\end{abstract}

\section*{INTRODUCTION}

The Phoswich Detection System (\pds) is one of the Narrow Field Instrument
aboard the X--ray astronomy satellite \B, a program of the Italian Space Agency
(ASI) with Dutch participation \cite{Boella97a}. A detailed description of the
instrument can be found in \cite{Frontera97}.

The \pds\ was designed to operate in the hard X--ray range from 15 to 300 keV
and to perform high sensitivity spectroscopic and temporal studies of celestial
X--ray sources, in particular their continuum emission. Classes of sources
accessible to \pds\ include High Mass X--ray Binaries (HMXRB), Low Mass X--ray
Binaries (LMXRB), Am Her--type sources, supernova remnants (in particular
Crab-like sources) and Active Galactic Nuclei (AGN).

In this paper we report on functional performance of \pds\ and on some
scientific results obtained from the observation of different classes of X--ray
sources performed during the Science Verification Phase (SVP) of the \B\
satellite.

\section*{IN-FLIGHT FUNCTIONAL PERFORMANCE}

\B\ was launched from Cape Canaveral with an Atlas-Centaur rocket on April 30,
1996. Its orbit is almost equatorial (3.9$^\circ$) at an altitude of 600 Km.
During the \B\ Commissioning Phase (1 May--30 June 1996), the \pds\ was
switched on and tested. Since then all subsystems continue to properly operate
with very good performance.

The anticoincidence (AC) shields provide a reduction of the background level by
about a factor two with respect to the level obtained with the phoswich
technique alone. In addition, the AC system strongly decreases the background
modulation along the \B\ orbit.

The background level $B$ of the \pds\ is the lowest obtained thus far with high
energy instruments at satellite orbits, specially at energies beyond 100 keV.
In the 15--300 keV band, $B$ is about $2.0\times 10^{-4}$ 
Cts~cm$^{-2}$~sec$^{-1}$~keV$^{-1}$: this is $\sim 70$\% of that published in
the \B\ handbook \cite{Piro95}. In 100--300 keV $B$ is a factor 3 lower than
that expected. The background modulation along a single \B\ orbit and on one
day time scale is about 20\%, while longer term variations ({\em e.g.\/},
build-up effects) are negligible.

We evaluated the systematic error in the background subtraction introduced by
the rocking collimator technique, by computing the background level variation
between the ON- and OFF-source positions in 32 ksec observing time. We obtained
$0.17\pm 0.09$ mCrab in the 15--300 keV energy band, corresponding to a
5$\sigma$ instrument sensitivity of 0.9 mCrab. This has to be compared to the
value of 0.5 mCrab, if only Poisson statistics is taken into account.

\section*{SCIENTIFIC RESULTS}

We will discuss some scientific results obtained with \pds\ during the \B\ SVP,
that show the actual spectral and timing capability of the instrument in
flight.

\subsection*{Crab Nebula}
 
The source has been observed for calibration purposes two times, in September
1996 (obs.I) and in April 1997 (obs.II). The spectral deconvolution makes use
of a response function derived from a Monte Carlo code that describes the
interaction of a photon beam with the \pds\ instrument, complemented by the
ground calibrations \cite{Zavattini97}.

By fitting the total spectrum of Crab with a single power law model we 
obtained a photon index $\alpha$ consistent for both observations:
$\alpha=2.119\pm 0.002$ for obs.I and $\alpha=2.112\pm 0.003$ for obs.II. The
normalization parameters at 1 keV are $8.60\pm 0.06$ and $8.57\pm 0.04$ for
obs.I and obs.II, respectively. The fitting was not so good: the reduced
$\chi^2$ was 2.89 (77 dof) for obs.I and 3.09 (67 dof) for obs.II. These high
$\chi^2$  are partly due to a break in the Crab high energy spectrum also
observed by other groups \cite{Bartlett94}, and partly to wiggles in the count
rate spectrum between 30 and 60 keV, whose instrumental origin is under
investigation. These systematic effects are estimated to be less than 5\% in
the residuals and less than 1\% in the spectral index reconstruction.

We can conclude, from these results, that the photon index is consistent with
previous results on the source spectrum \cite{Bartlett94}, while the
normalization parameter is about 10\% lower than the extrapolation of the power
law spectrum measured with the MECS instrument \cite{Boella97} aboard \B. It is
however consistent with published results on Crab \cite{Bartlett94} within
their uncertainties.

\subsection*{Centaurus X--3}

\begin{figure}
\epsfxsize=8cm
\centerline{\epsffile{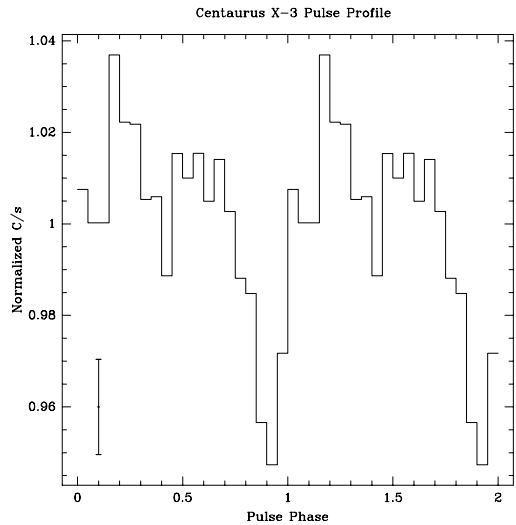}}
\caption{\B/\pds\ pulse profile of the 4.84 sec X--ray binary pulsar Cen X--3.}
\label{cenx3}
\end{figure}

This classical X--ray pulsar, belonging to the class of HMXRBs, was observed
for $\sim 6000$ sec during a low intensity level (30 mCrab) at high energies.
The source was clearly detected up to about 50 keV. The hard X--ray spectrum of
the source is consistent with an optically thin thermal bremsstrahlung with
$kT=8.2\pm 0.2$ keV and normalization parameter $0.66\pm 0.03$. The plasma
temperature and flux level are consistent with those measured by HEAO--1/A4
\cite{Howe83}. The \pds\ pulse profile of the pulsar ($P_P\sim 4.84$ sec) is
characterized by a dip (see Fig.~\ref{cenx3}). It appears different from the
average pulse profiles previously reported \cite{Howe83}. An investigation on
the origin of this different behavior is under way.

\subsection*{4U1626--67}
 
\begin{figure}
\centerline{\epsfxsize=0.55\textwidth\epsffile{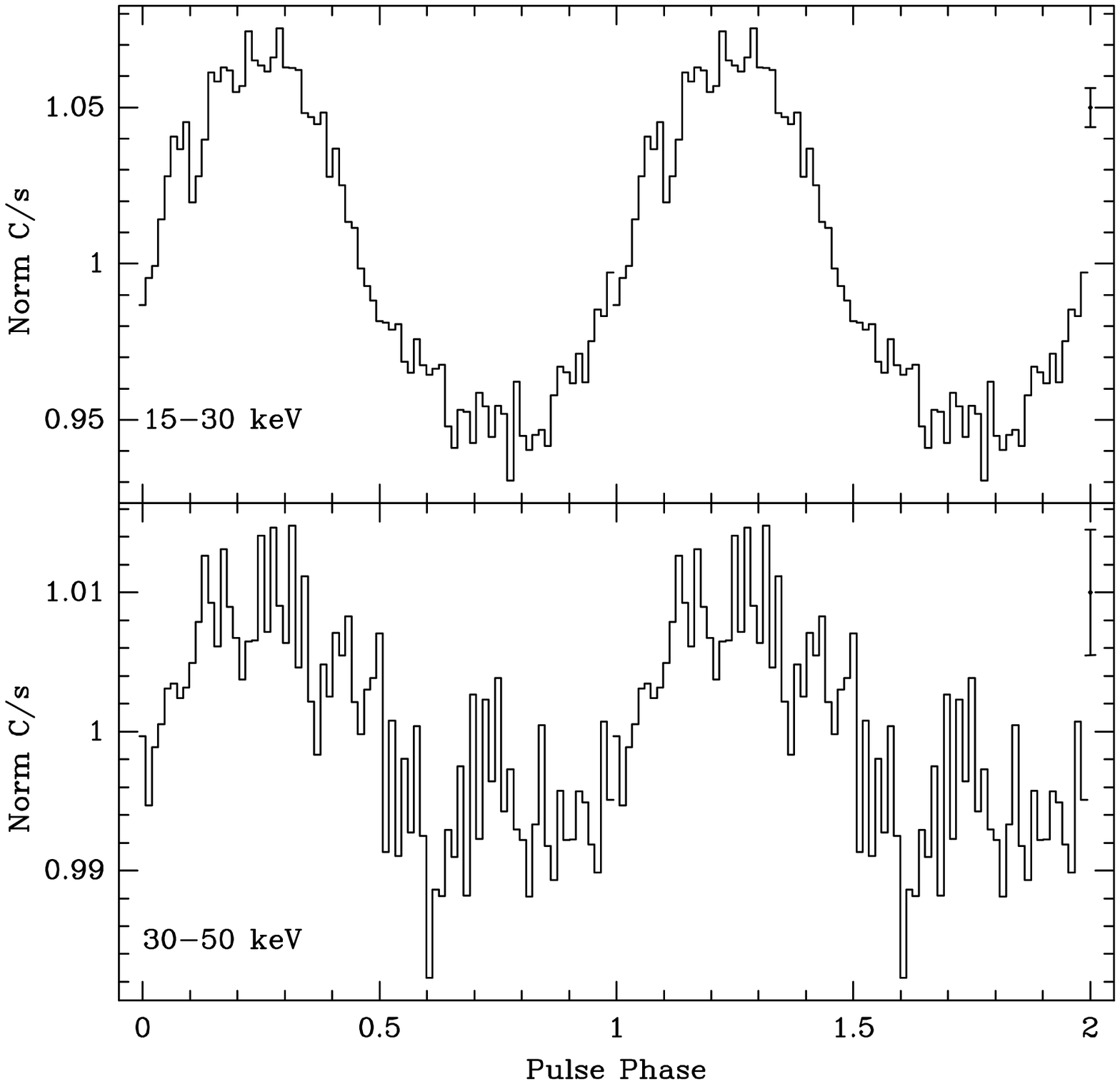}
\epsfxsize=0.55\textwidth\epsffile{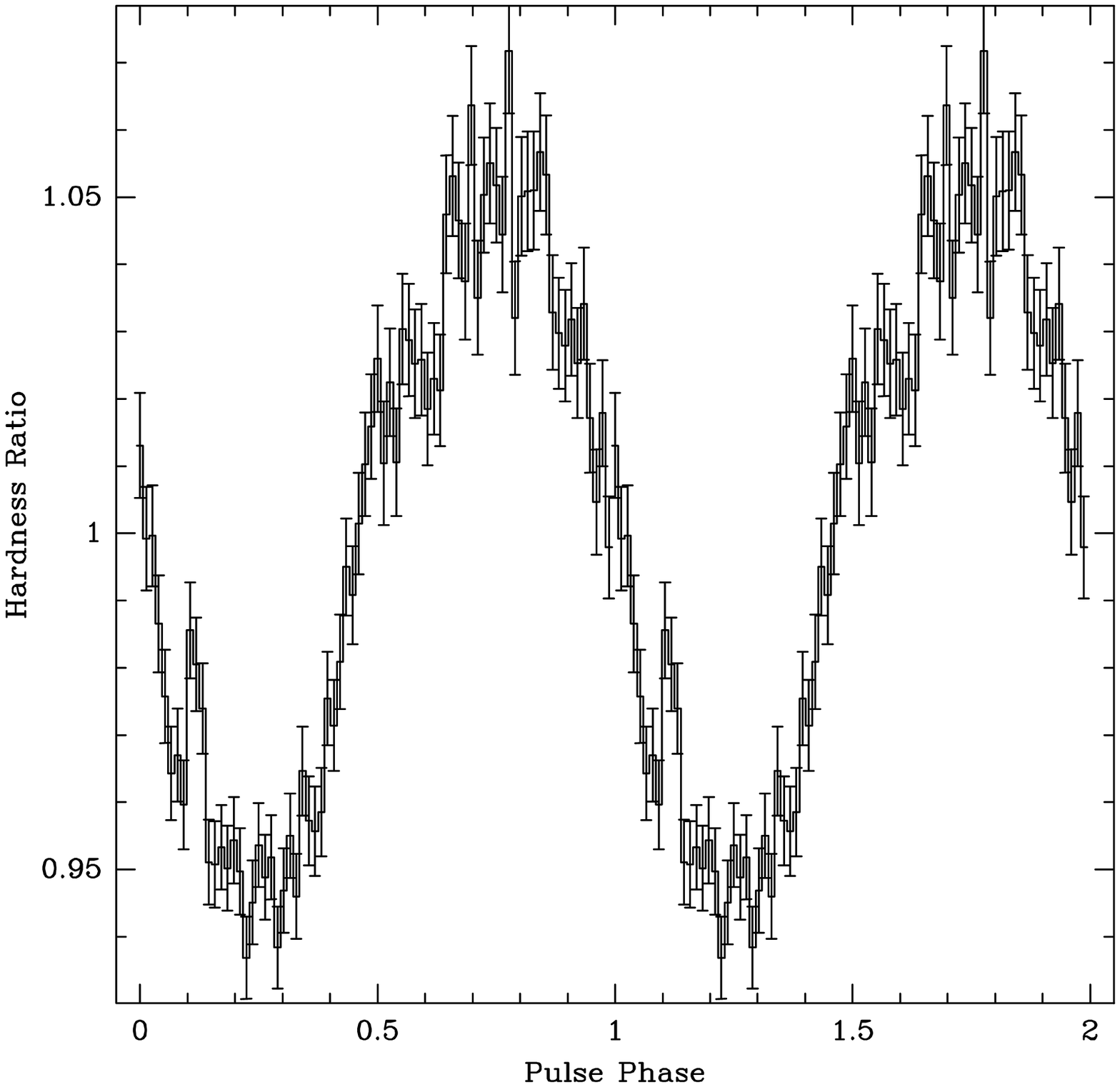}}
\caption[]{a): 4U1626--67 pulse profiles in two energy bands as observed by
\pds. b): Hardness ratio between the 30--50 and 15--30 keV pulse profiles.}
\label{1626}
\end{figure}

This X--ray pulsar ($P_P\sim 7.7$ sec) is one of the few LMXRBs that show
pulsed emission. It was detected with \pds\ up to 50 keV in $\sim 70$ ksec
exposure time. The spectral analysis is in progress, while the pulse profiles
of the pulsar in two hard X--ray ranges are shown in Fig.~\ref{1626}a. As can
be seen, 4U1626--67 exhibits a sinusoidal shape and a peculiar feature: the
hardness ratio between 30--50 keV and 10--30 keV pulse profiles turns out to be
anti-correlated with the intensity profile (Fig.~\ref{1626}b).

\subsection*{PKS2155--304}

\begin{figure}
\epsfysize=9cm
\centerline{\epsffile{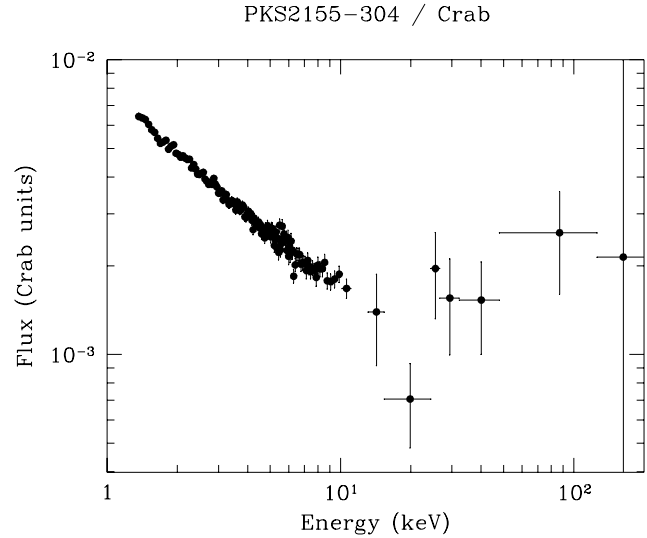}}
\caption[]{Observed ratio between PKS2155--305 and Crab spectra. Data from both
MECS and \pds\ were used.}
\label{pks}
\end{figure}

PKS2155--304 is one of the strongest BL Lac objects in the 2--10 keV energy
band. It is the first time the source has been simultaneously observed in a
broad-energy band (0.1--300 keV), crucial for studying the relationship among
different emission components. The source was clearly detected by \pds\ up to
100 keV (exposure time $\sim 100$ ksec). The ratio between the source spectrum
and the Crab spectrum is shown in Fig.~\ref{pks}. It is apparent that the
source spectrum is softer than Crab below 10--20 keV, while it is harder or
similar to the Crab spectral slope above 10--20 keV. An extended paper on the
\B\ results on this source can be found elsewhere \cite{Giommi97}.

The most probable interpretation of the hard X--ray component is Inverse 
Compton radiation associated to a Self-Synchro Compton (SSC) process, while the
lower energy component is synchrotron radiation.

\subsection*{CONCLUSION}

The SAX/PDS experiment is performing as designed and shows a flux sensitivity
that is in agreement or better than that given in the \B\ handbook
\cite{Piro95}.

\end{document}